\documentclass[12pt]{article}

\pdfoutput=1

\usepackage[bulletsep]{collref}
\usepackage{amssymb,graphicx}
\usepackage[intlimits]{amsmath}
\usepackage{bbm}
\usepackage[small]{subfigure}

\usepackage{MnSymbol}

\makeatletter \@addtoreset{equation}{section} \makeatother

\makeatletter
\let\old@startsection=\@startsection
\let\oldl@section=\l@section
\renewcommand{\@startsection}[6]{\old@startsection{#1}{#2}{#3}{#4}{#5}{#6\mathversion{bold}}}
\renewcommand{\l@section}[2]{\oldl@section{\mathversion{bold}#1}{#2}}
\makeatother

\makeatletter
\let\old@makecaption=\@makecaption
\def\@makecaption{\small\old@makecaption}
\makeatother

\newcommand{\be}{\begin{equation}}\newcommand{\ee}{\end{equation}}
\newcommand{\bea}{\begin{eqnarray}} \newcommand{\eea}{\end{eqnarray}}

\begin{document}

\setcounter{page}{1}
\renewcommand{\thefootnote}{\arabic{footnote}}
\setcounter{footnote}{0}

\thispagestyle{empty}
\begin{flushright}\footnotesize
\texttt{NORDITA-2017-135} \\
\texttt{UUITP-50/17}
\vspace{0.6cm}
\end{flushright}

\renewcommand{\thefootnote}{\fnsymbol{footnote}}
\setcounter{footnote}{0}

\mbox{}
\vspace{0truecm}
\linespread{1.1}

\begin{center}
{\Large\textbf{\mathversion{bold} Wilson loops in antisymmetric representations \\ from localization in supersymmetric gauge theories
}
\par}

\vspace{0.8cm}

\textrm{Jorge~G.~Russo$^{1,2}$  and Konstantin~Zarembo$^{3,4}$\footnote{Also at ITEP, Moscow, Russia}}
\vspace{4mm}

\textit{${}^1$ Instituci\'o Catalana de Recerca i Estudis Avan\c cats (ICREA), \\
Pg. Lluis Companys, 23, 08010 Barcelona, Spain.}\\
\textit{${}^2$ Departament de F\' \i sica Cu\' antica i Astrof\'\i sica and ICCUB\\
Universitat de Barcelona, Mart\'i Franqu\`es, 1, 08028
Barcelona, Spain. }\\
\medskip
\textit{${}^3$Nordita, KTH Royal Institute of Technology and Stockholm University,
Roslagstullsbacken 23, SE-106 91 Stockholm, Sweden.}\\
\textit{${}^4$Department of Physics and Astronomy, Uppsala University\\
SE-751 08 Uppsala, Sweden.}\\
\vspace{0.2cm}
\texttt{jorge.russo@icrea.cat, zarembo@nordita.org}

\vspace{5mm}


\begin{flushright}
{\it \large In memory of Ludvig Dmitrievich Faddeev}
\end{flushright}

\par\vspace{0.4cm}

\textbf{Abstract} \vspace{3mm}

\begin{minipage}{13cm}

Large-$N$ phase transitions occurring in massive ${\cal N}=2$ theories can be probed by Wilson loops in large antisymmetric representations. The logarithm of the Wilson loop is effectively described by the free energy of a Fermi distribution and exhibits second-order phase transitions (discontinuities in the second derivatives) as the size of representation varies. We illustrate the general features of antisymmetric Wilson loops on a number of examples where the phase transitions are known to occur: ${\cal N}=2 $ SQCD with various mass arrangements and $\mathcal{N}=2^*$ theory. As a byproduct we solve planar ${\cal N}=2$ SQCD  with three independent mass parameters. This model has two effective mass scales and   undergoes two phase transitions.
\end{minipage}

\end{center}

\newpage
\setcounter{page}{1}
\renewcommand{\thefootnote}{\arabic{footnote}}
\setcounter{footnote}{0}

\tableofcontents

\section{Introduction}

Localization is a powerful tool to explore supersymmetric gauge theories in the non-perturbative domain, and in particular in the large-$N$ limit. Exact results obtained in the latter case bear direct links to the holographic duality at strong coupling. 

The partition function and select observables of any $\mathcal{N}=2$ gauge theory on $S^4$ localize to an effective matrix model \cite{Pestun:2007rz}, that can be studied at large-$N$ by the standard methods of random matrix theory \cite{Brezin:1977sv}. A somewhat unexpected outcome of this analysis is appearance of large-$N$ phase transitions in a variety of massive $\mathcal{N}=2$ gauge theories \cite{Russo:2013qaa,Russo:2013kea}. 
Their holographic description remains an interesting open problem. The difficulty is related to the strong-weak coupling nature of the holographic duality: transitions occur upon varying a coupling constant, and this is difficult to achieve in holography where an infinite-coupling regime is normally considered. It is desirable, in this respect, to devise observables that undergo phase transitions at fixed coupling while some other auxiliary parameter is being varied.
Particularly promising probes of this kind are Wilson loops in large antisymmetric representations of the gauge group, which were shown to undergo phase transitions once the size of representation is dialed to a critical value \cite{Chen-Lin:2015dfa}. 

Methods to calculate expectation values of Wilson loops in large representations, both holographically and from localization, have been devised in the context of the $\mathcal{N}=4$ super-Yang-Mills (SYM) theory  \cite{Drukker:2005kx,Yamaguchi:2006tq,Gomis:2006sb,RodriguezGomez:2006zz,Hartnoll:2006is,Gomis:2006im,Yamaguchi:2007ps,Okuda:2008px,Faraggi:2011ge,Fiol:2013hna,Buchbinder:2014nia,Faraggi:2014tna,Fiol:2014vqa,Horikoshi:2016hds,Zarembo:2016bbk,Chen-Lin:2016kkk,Gordon:2017dvy}, which localizes to the Gaussian matrix model \cite{Erickson:2000af,Drukker:2000rr,Pestun:2007rz}. These methods have been transplanted to $\mathcal{N}=2$ theories, both conformal \cite{Fraser:2011qa,Fraser:2015xha} and  massive \cite{Chen-Lin:2015dfa,Liu:2017fiq}, and in massive case antisymmetric Wilson loops were shown to undergo phase transitions in the rank of the gauge-group representation \cite{Chen-Lin:2015dfa}.

 Here we discuss general features of phase transitions in antisymmetric Wilson loops and then apply the general framework to a variety of theories where the transitions are known to occur, namely to $\mathcal{N}=2$ super-QCD for different combinations of mass parameters and to $\mathcal{N}=2^*$ SYM, extending in the latter case the results of \cite{Chen-Lin:2015dfa,Liu:2017fiq}.

\section{Anti-symmetric Wilson loops}

By localization, the expectation value of the circular Wilson loop in any $\mathcal{N}=2$ theory on $S^4$ can be expressed as a matrix model correlator:
\begin{equation}\label{WilsC}
 W_{\mathcal{R}}(C)=\left\langle \mathop{\mathrm{tr}}\nolimits_{\mathcal{R}}\,{\rm e}\,^{L\Phi }\right\rangle_{\rm MM},
\end{equation}
where $L=2\pi R_{S^4}$ is the circumference of the circle, which we consider to be large compared to any other scale in the problem.  In the decompactification limit $R_{S^4}\rightarrow \infty $ the circular loop should obey the same universal scaling law as any sufficiently large contour. We thus expect that (\ref{WilsC}) describes the universal behavior of large Wilson loops of arbitrary shape in $\mathcal{N}=2$ gauge theories.

We concentrate on the Wilson loops in antisymmetric representations. The generating function for characters in the rank-$k$ anti-symmetric representation $\mathcal{A}_k$  is given by 
\begin{equation}
 \sum_{k=0}^{N}\,{\rm e}\,^{-L\nu k}
 \mathop{\mathrm{tr}}\nolimits_{\mathcal{A}_k}\,{\rm e}\,^{L\Phi } 
 =\det\left(1+\,{\rm e}\,^{L(\Phi -\nu )}\right).
\end{equation}
An expectation value in any particular representation can  be computed as
\begin{equation}\label{intrep}
 W_{\mathcal{A}_k}(C)=L \int_{C-\frac{i\pi }{L}}^{C+\frac{i\pi }{L}}\frac{d\nu }{2\pi i}\,\,\,{\rm e}\,^{Lk\nu }\left\langle\det\left(1+\,{\rm e}\,^{L(\Phi -\nu )}\right)\right\rangle_{\rm MM}.
\end{equation}

In the 't~Hooft limit, with
\begin{equation}
 N\rightarrow \infty ,\qquad k\rightarrow \infty ,\qquad f=\frac{k}{N}\,-\,{\rm fixed},
\end{equation}
the saddle point of the matrix model is not affected by the Wilson loop insertion, and is characterized by the eigenvalue density
\begin{equation}
 \rho (x)=\frac{1}{N}\,\mathop{\mathrm{tr}}\delta (x-\Phi ),
\end{equation}
obtained by solving the localization matrix model in the large-$N$ limit.
In the same scaling limit, the saddle-point approximation for the integral over $\nu $ in (\ref{intrep}) becomes exact and the Wilson loop expectation value takes on an exponential, perimeter-law form:
\begin{equation}
\label{wilAA}
 W_{\mathcal{A}_k}(C)\simeq \,{\rm e}\,^{NLF(f)}.
\end{equation}

The function $F(f)$ is determined by the saddle-point equation, which can written as follows. Define the function
\begin{equation}
 \mathcal{F}(f,\nu )=f\nu +\frac{1}{L}\int dx\,\rho (x)\ln\left(1+\,{\rm e}\,^{L(x-\nu )}\right).
\end{equation}
Then $F(f)=\mathcal{F}(f,\nu (f))$, where $\nu (f)$ is the value of $\nu $ that maximizes $\mathcal{F}(f,\nu )$ for a given $f$:
\begin{equation}
 f=\int_{}^{}\frac{dx\,\rho (x)}{\,{\rm e}\,^{L(x-\nu(f) )}+1}\,.
\end{equation}
Once the eigenvalue distribution is known, the expectation value of the antisymmetric Wilson loop can be computed from the above two equations \cite{Yamaguchi:2006tq}.

The distribution of eigenvalues in massive $\mathcal{N}=2$ theories is confined to an interval $(-\mu ,\mu )$, where $\mu $ is the characteristic mass scale of the underlying gauge theory, which we assume to be such that $\mu L\gg 1$. This corresponds to the low-temperature regime of the effective statistical model, and the Fermi distribution in the above formulas can be replaced by the step function:
\begin{eqnarray}\label{statmech}
 \mathcal{F}(f,\nu )&=&f\nu +\int_{\nu }^{\mu }dx\,\rho (x)\left(x-\nu \right)
\nonumber \\
f&=&\int_{\nu(f) }^{\mu }dx\,\rho (x).
\end{eqnarray}
It follows from these equations that
\begin{eqnarray}\label{derivatives of F}
 \frac{dF}{df}&=&\nu 
\nonumber \\
\frac{d^2F}{df^2}&=&-\frac{1}{\rho (\nu )}\,,
\end{eqnarray}
where $\nu \equiv \nu (f)$.
While the first of these equations is exact, the second one is only valid in the limit $\mu L\rightarrow \infty $.

\subsection{Phase transitions}

The eigenvalue distribution in massive theories may develop specific singularities in a certain range of parameters, which lead to phase transitions as the parameters change. The fundamental Wilson loops are not very good probes of the phase transitions, because the influence of any given singularity in the eigenvalue density is washed out by averaging over the whole eigenvalue distribution. On the contrary, the singularities are very pronounced in  large antisymmetric Wilson loops due to the sharp form of the Fermi distribution at zero temperature. 

Large-$N$ solutions of the localization matrix model are known for a number of $D=4$ supersymmetric gauge theories \cite{Russo:2013kea}. It was observed that in the decompactification limit, when the size of the four-sphere becomes infinite, $R_{S^4}\rightarrow \infty $, the density may develop singularities.
The features encountered so far are come in two types, the delta functions and the one-sided cusps. The first type of singularity arises in theories with fundamental matter, such as super-QCD, while the second type is characteristic for adjoint matter and arises, for instance, in the $\mathcal{N}=2^*$ theory. It is these singularities which are responsible for the phase transitions.
We consider the two cases in turn, first at a general, model-independent level and then on concrete examples.

\subsubsection{Delta-function singularity}
\label{deltasingu}

Suppose that the density has a delta-functional peak in the middle of the eigenvalue distribution:
\begin{equation}
 \rho (x)\stackrel{x\rightarrow x_c}{=}\rho_0+p\delta \left(x-x_c\right),
\end{equation}
where $p$ is the fraction of eigenvalues concentrated in the peak and $\rho _0$ is a constant. It is not hard to see that this structure translates into two singularities in the Wilson loop expectation value, at $f=f_{c_+}$ and $f=f_{c_-}$, where
\begin{equation}
 f_{c_+}=\int_{x_c+0}^{\mu }dx\,\rho (x), \qquad 
 f_{c_-}=f_{c_+}+p.
\end{equation}
The function $\nu (f)$ stays flat, $\nu (f)=x_c$, for $f_{c_+}<f<f_{c_-}$. From (\ref{derivatives of F}) we see that the free energy $F$ is continuous across the transitions together with its first derivative, while the second derivative experiences a finite jump:
\begin{equation}\label{jump}
 \left.\frac{d^2F}{df^2}\right|_{f_{c_\pm}-0}^{f_{c_\pm}+0}=\pm\frac{1}{\rho _0}\,.
\end{equation}

\subsubsection{Cusp}

Strictly speaking, there are two types of cusps, the left cusp and the right cusp. On one side of the cusp the density approaches a finite value, while on the other side it has an inverse square root singularity:
\begin{equation}
 \rho (x)\stackrel{x\rightarrow x_{c_\pm}}{=}
\begin{cases}
 \frac{C}{\sqrt{|x-x_{c_\pm}|}} & {\rm at}~x\rightarrow x_{c_\pm}\mp 0
\\
  \rho _0 & {\rm at}~x\rightarrow x_{c_\pm}\pm 0.
\end{cases}
\end{equation}
As a result, the antisymmetric Wilson loop develops a singularity at $f=f_{c_\pm}$, where
\begin{equation}
 f_{c_\pm}=\int_{x_{c_\pm}}^{\mu }dx\,\rho (x).
\end{equation}
The free energy stays continuous together with its first derivative across the critical point, while the second derivative experiences a finite jump, given by the same formula (\ref{jump}) as for the delta-function singularity.

These are the two types of singularities encountered in the large-$N$ solutions of the localization matrix models in $\mathcal{N}=2$ theories in four dimensions. The delta-function singularities arise in theories fundamental matter, while cusps are characteristic of the adjoint matter. Below we consider a few examples of each type. 

\section{Pure ${\cal N}=2$ SYM theory}

We begin with a case where there is no large $N$ phase transition,
${\cal N}=2$ $SU(N)$ SYM theory without matter.
The large-$N$ solution of this theory was first obtained from the 
Seiberg-Witten theory \cite{Douglas:1995nw}.
The localization matrix model for pure $\mathcal{N}=2$ SYM was studied in \cite{Russo:2012ay,Russo:2013sba}.
In the decompactification limit $R\to\infty $, one finds 
\cite{Douglas:1995nw,Russo:2012ay}:
\be
\rho(x) = \frac{1}{\pi \sqrt{\mu^2-x^2}}\ ,\qquad \mu =c_0\Lambda R\ ,\quad c_0 =2 e^{-1-\gamma}\ .
\ee
where $R$ is the radius of $S^4$ and $\Lambda $ is the dynamically generated scale.\footnote{In terms of the original, renormalized 't Hooft coupling, $\Lambda R = \exp[-4\pi^2/\lambda] $.}

The saddle-point equation (\ref{statmech}) gives
\be
f= \int_{\nu}^\mu dx\  \frac{1}{\pi \sqrt{\mu^2-x^2}}=\frac{1}{\pi}{\rm arccos}\frac{\nu}{\mu}\ ,
\ee
{\it i.e.}
\be
\nu = \mu \cos(f\pi )\ .
\ee
The Wilson loop (\ref{wilAA}) is given by
\be
\ln W_{\mathcal{A}_k} = N \int_{\nu}^\mu dx\ \rho (x)\ x =\frac{N}{\pi} \sqrt{\mu^2-\nu^2}\ .
\ee
Thus
\be
\ln W_{\mathcal{A}_k}  = NRF(f)\ ,\qquad 
F(f) = \frac{c_0\Lambda}{\pi}  \sin (f\pi )\ ,\qquad 0<f<1\ .
\ee
It is a smooth function of $f$, as expected,
since the theory does not have any phase transitions.

\section{$\mathcal{N}=2$ Super QCD} \label{sectionSQCD}


As a first example where phase transitions occur, we consider the case of $\mathcal{N}=2$ $SU(N)$
super Yang-Mills theory with $N_f$ pairs of fundamental and anti-fundamental hypermultiplets of masses $(M, -M)$, $N_f<N$.\footnote{The theory with $N_f=N$ with massless hypermultiplets
corresponds to $\mathcal{N}=2$ superconformal SQCD.
In this case there are no phase transitions \cite{Russo:2013kea}.}

This model was studied in \cite{Russo:2013kea,Russo:2013sba}.
We consider the  $N\to \infty$ limit
with fixed Veneziano parameter $\zeta \equiv N_f/N$ and fixed (renormalized) 't Hooft coupling $\lambda = g_{\rm YM}^2 N$, which in turn is traded by the dynamically generated scale $\Lambda $,
\be
\Lambda R= e^{-\frac{4\pi^2}{\lambda (1-\zeta)}}
\ee
In the decompactification limit, the theory undergoes a phase transition at $\Lambda =M/2$. The different features of the model were 
also reproduced in \cite{Russo:2015vva} by taking the large $N$ limit on the the Seiberg-Witten curve. 
Another interesting limit is the decompactification limit  at finite $N$. 
The partition function can then be computed by incorporating instantons through the Seiberg-Witten curve  \cite{Russo:2014nka}.

It is possible to generalize
the theory by considering fundamental hypermultiplets of different
masses, $M_1,...,M_{N_f}$. The eigenvalue density in the large $N$ limit can still be determined in terms of
analytic formulas.
 The resulting theory describes multiple phase transitions occurring whenever the largest eigenvalue $\mu$ (or the lowest eigenvalue $-\mu $) crosses a new mass scale. An explicit example with
three  mass scales, $ \Lambda$, $m, M$ is given in the appendix~\ref{scales-appx}.

The present theory depends on
two parameters $\Lambda R$ and $M R$. In the
large $N$ limit, the eigenvalue density is determined by a saddle-point equation.
This simplifies in the decompactification limit $R\to\infty $. By differentiating the saddle-point equation once, one obtains, for $R\to\infty$,
\be\label{auxiliary-SQCD}
2\int_{-\mu }^{\mu } dy\,\rho(y)\ln \frac{\left(x-y\right)^2}{\Lambda ^2}
= \zeta \ln\frac{\left(x^2-M^2\right)^2}{\Lambda ^4}\,.
\ee
Differentiating once more, we find the equation:
\be
\label{mfree}
2\strokedint_{-\mu }^{\mu } dy\,\,  \frac{\rho(y)}{x-y}
=   \frac{\zeta }{x+M} +\frac{\zeta }{x-M}\,.
\ee
This (singular) integral equation has two different solutions, which depends on whether the points  $x=\pm M$  lie, or do not lie, within the eigenvalue distribution. 
The two solutions describe the two phases of the theory:

\begin{itemize}

\item Phase I. Strong coupling phase, $\mu >M$, with

\be
\rho(x) = \frac{1-\zeta}{\pi \sqrt{\mu^2-x^2}}+\frac{\zeta}{2} \delta(x+M)+\frac{\zeta}{2} \delta(x-M)
\ee

\item Phase II. Weak coupling phase, $\mu <M$, with

\be
\rho(x) = \frac{1-\zeta}{\pi \sqrt{\mu^2-x^2}}+\zeta M\sqrt{M^2-\mu^2}\frac{1}{\pi \sqrt{\mu^2-x^2} (M^2-x^2)}
\ee

\end{itemize}

The non-analytic behavior in the Wilson loop $\ln W_{\mathcal{A}_k}$ appears when $\nu(f)$ crosses a singular point in the eigenvalue density.
Since in phase II the eigenvalue density is smooth, in the weak coupling phase the $\ln W_{\mathcal{A}_k}$ will be smooth.
We thus focus on the more interesting strong coupling phase, where we have a delta-function singularity of the type described in section \ref{deltasingu}.
In this phase, $\mu =2\Lambda $, so the phase transition takes place at $\Lambda_c =M/2$ (see appendix~\ref{scales-appx}).

The saddle-point equation (\ref{statmech}) now  leads to
\be
f= (1-\zeta)\frac{1}{\pi}{\rm arccos}\frac{\nu}{\mu}+\frac{\zeta}{2} \theta(M-\nu)+\frac{\zeta}{2} \theta(-M-\nu)\ .
\ee
The solution has several critical points $f_1,\ f_2,\ f_3, \ f_4$, where
\be
\pi f_1=(1-\zeta)\, {\rm arccos}\frac{M}{\mu}\ ,\quad f_2 =f_1+\frac{\zeta}{2} \ ,\quad f_3=1-f_2\ ,\quad f_4=1-f_1\ .
\ee
The critical points $(f_1,f_2)$ corresponds
to the points $(f_{c_+},f_{c_-})$ described in section 2.1.1, when $\nu $ meets the delta-function singularity at $x =M$. In the interval $(f_1,f_2)$,
$\nu $ remains constant, $\nu =M$.
Similarly, $(f_3,f_4)$ are
the two critical points corresponding to the delta-function singularity at $x=-M$. 
Explicitly, the complete solution is 
\bea
&& \nu =\mu \cos \frac{\pi f}{1-\zeta}\ ,\qquad 0<f<f_1
\nonumber\\
&& \nu =M\ ,\qquad \qquad f_1<f<f_2
\nonumber\\
&& \nu =\mu \cos \frac{\pi(f-\zeta/2) }{1-\zeta}\ ,\qquad f_3<f<f_2
\nonumber\\
&& \nu =-M\ ,\qquad \quad f_3<f<f_4
\nonumber\\
&& \nu =\mu \cos \frac{\pi (f-\zeta)}{1-\zeta}\ ,\qquad f_4<f<1
\nonumber\\
\eea
Fig. 1 shows a plot of the saddle-point $\nu $ as a function of $f=k/N$ for $\zeta=1/2$ ({\it i.e.} $N_f=N/2$).

\begin{figure}[ht!]
\centering
\includegraphics[width=0.6\textwidth]{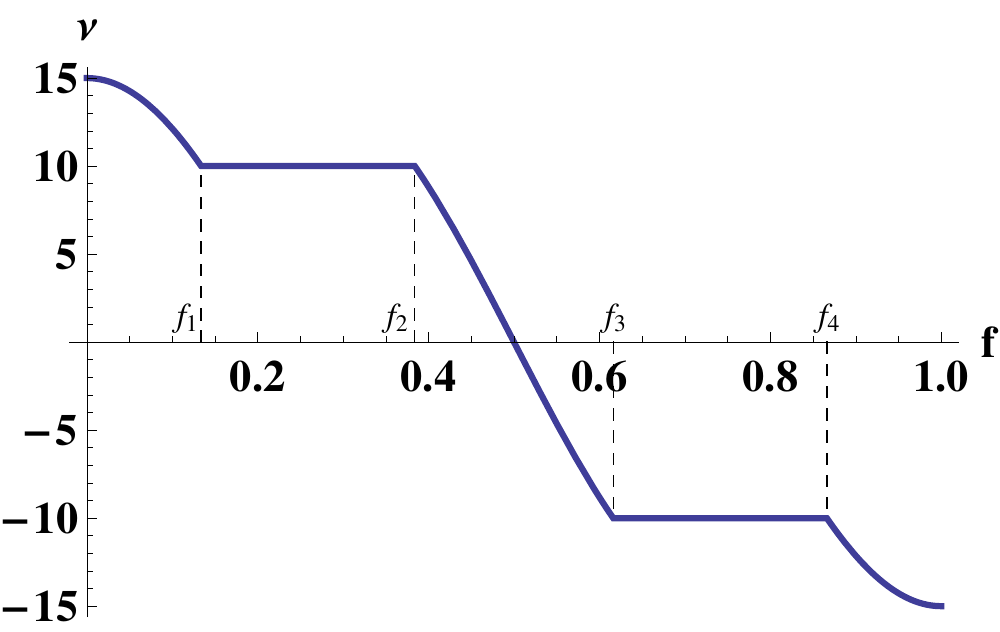}
\caption{$\nu $ as a function of $f$ ($M=10,\ \Lambda=3M/4,\ \zeta=1/2$). As $\nu =dF/df$, the figure also shows the behavior of the first derivative of 
$\ln W_{\mathcal{A}_k}$.}
\label{nuf}
\end{figure}

\smallskip

Let us now compute the Wilson loop. We have $\ln W_{\mathcal{A}_k}=NR F(f)$, with
\bea
F  &=&  f \nu+  \int_{\nu}^\mu dx\ \rho (x)\ (x-\nu) 
\nonumber\\
&=&  f \nu+ \frac{1}{\pi} (1-\zeta ) \sqrt{\mu^2-\nu^2}- \frac{\nu}{\pi}(1-\zeta) {\rm arccos}\frac{\nu}{\mu}
\nonumber\\
&+& \frac{\zeta }{2}(M-\nu )\, \theta(M-\nu)
-\frac{\zeta }{2}(M+\nu)\, \theta(-M-\nu)\ .
\eea
It exhibits discontinuities in the second derivative with respect to $f$ at the critical points.
This is readily seen  from 
(\ref{derivatives of F}), since the first derivative of $ \ln W_{\mathcal{A}_k} $ is proportional to $\nu(f)$ which, as can be seen from fig. 1, has  discontinuities in its first derivatives.
{}More directly, from (\ref{jump}), we have
\be
\left.\frac{d^2F}{df^2}\right|_{f_{c_\pm}-0}^{f_{c_\pm}+0}=
\pm \frac{\pi\sqrt{\mu^2-M^2}}{1-\zeta}\, ,
\label{salto}
\ee
where  $f_{c_+},\ f_{c_-}$ are either $f_1,\ f_2$ or
$f_3,\ f_4$. 
Figs. 2a and 2b respectively show  plots of  $F=\frac{1}{NR}\ln W_{\mathcal{A}_k} $ and $d^2F/df^2$ as  functions of $f$. The first derivative, $dF/df$, being equal to $\nu(f)$, can be seen from figure 1. One can check that the jumps in the second derivatives in figure 2b
are in precise agreement with (\ref{salto}).

\begin{figure}[h!]
\centering
\begin{tabular}{cc}
\includegraphics[width=0.45\textwidth]{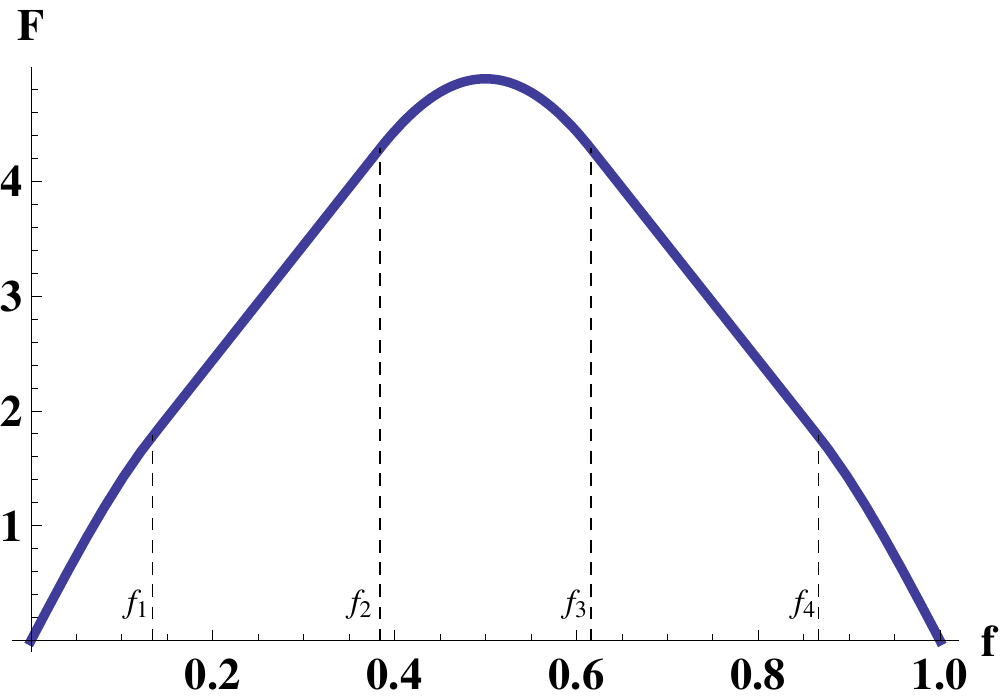}
&
\includegraphics[width=0.45\textwidth]{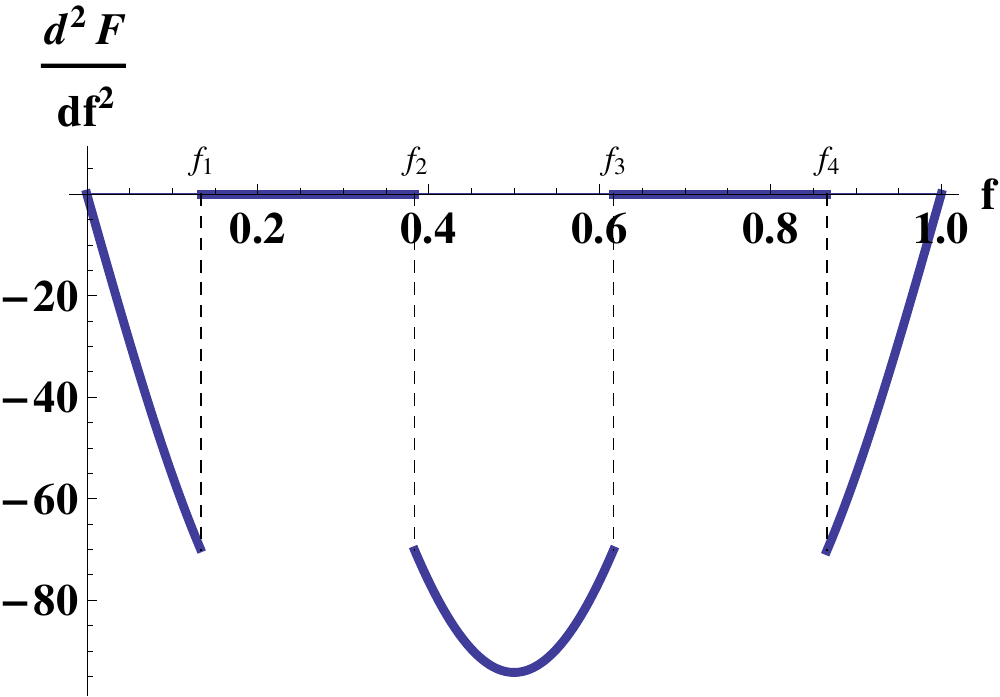}
\\
(a)&(b)
\end{tabular}
\caption{a) $F=\frac{1}{NR}\ln W_k $ as a function of $f$.  b) $\frac{d^2F}{df^2} $ as a function of $f$. Here $R=1$, with the same values for $M, \Lambda, \ \zeta$ as in fig. 1.}
\label{ddFf}
\end{figure}

\section{$\mathcal{N}=2^*$ theory}

The field content of $\mathcal{N}=2^*$ SYM consists of the vector multiplet and an adjoint hypermultiplet of mass $M$. The 't~Hooft coupling $\lambda =g^2_{\rm YM}N$ does not run and is just a parameter that characterizes the theory. The theory undergoes a fourth-order phase transition each time the width of the eigenvalue distribution $2\mu $ passes an integer multiple of $M$: $\mu(\lambda _c^{(n)})=nM/2 $. It is convenient to introduce the parameters
\begin{equation}
 n=\left[\frac{2\mu }{M}\right],\qquad \Delta =\left\{\frac{2\mu }{M}\right\},
\end{equation}
where the square (curly) brackets denote the integer (fractional) part of a real number. The eigenvalue density has left (right) cusps at
\begin{eqnarray}\label{SingN2*}
  x^{(k)}_{c_+}&=&\mu -kM,\qquad k=1,\ldots, n \nonumber \\
  x^{(k)}_{c_-}&=&\mu -kM-\Delta,\qquad k=0,\ldots, n-1. 
\end{eqnarray}
In what follows we concentrate on the strong-coupling regime of large $\lambda $.

\begin{figure}[t]
\begin{center}
\subfigure[]{
   \includegraphics[width=8.1cm] {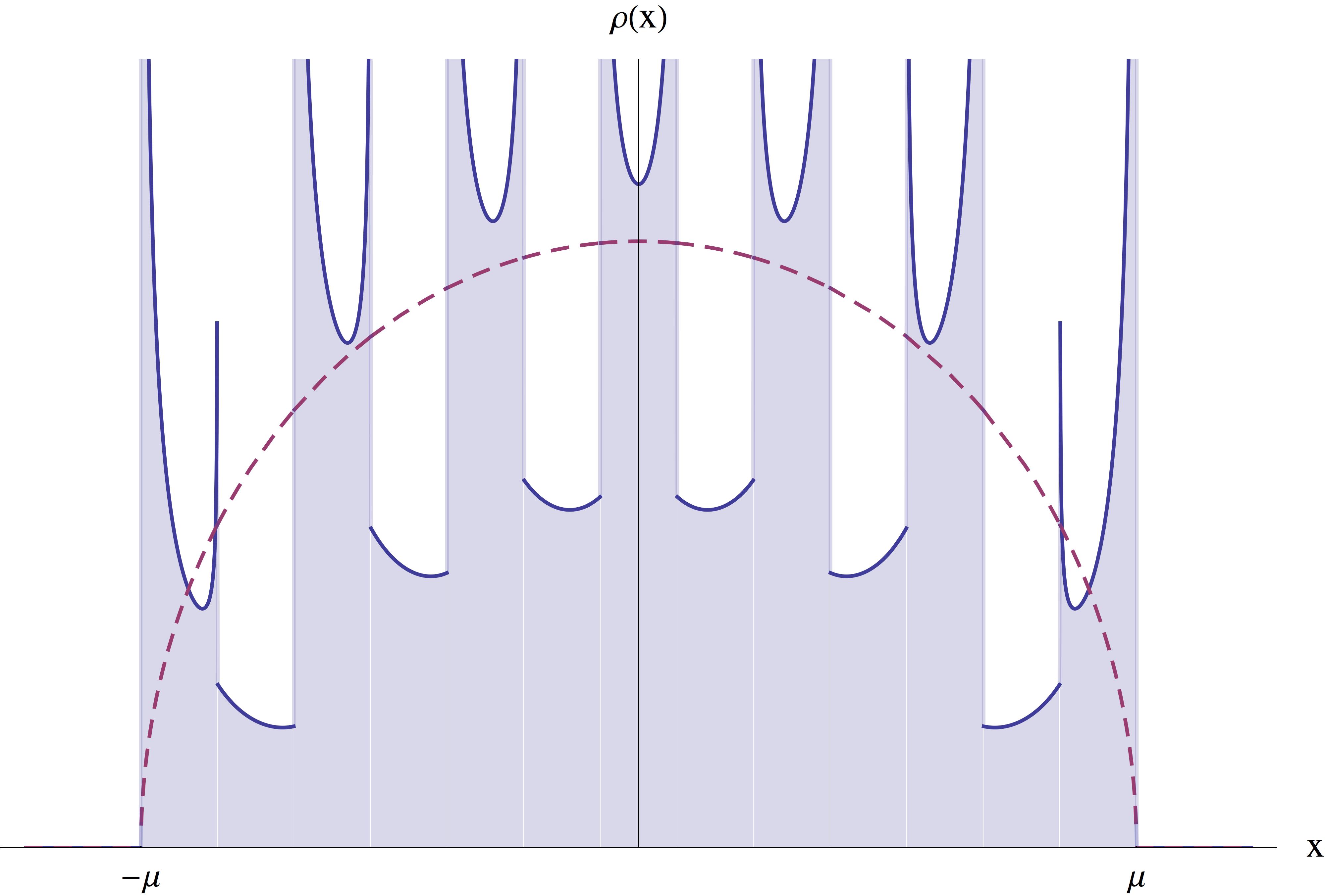}
   \label{fig:subfig1}
 }
 \subfigure[]{
   \includegraphics[width=8.1cm] {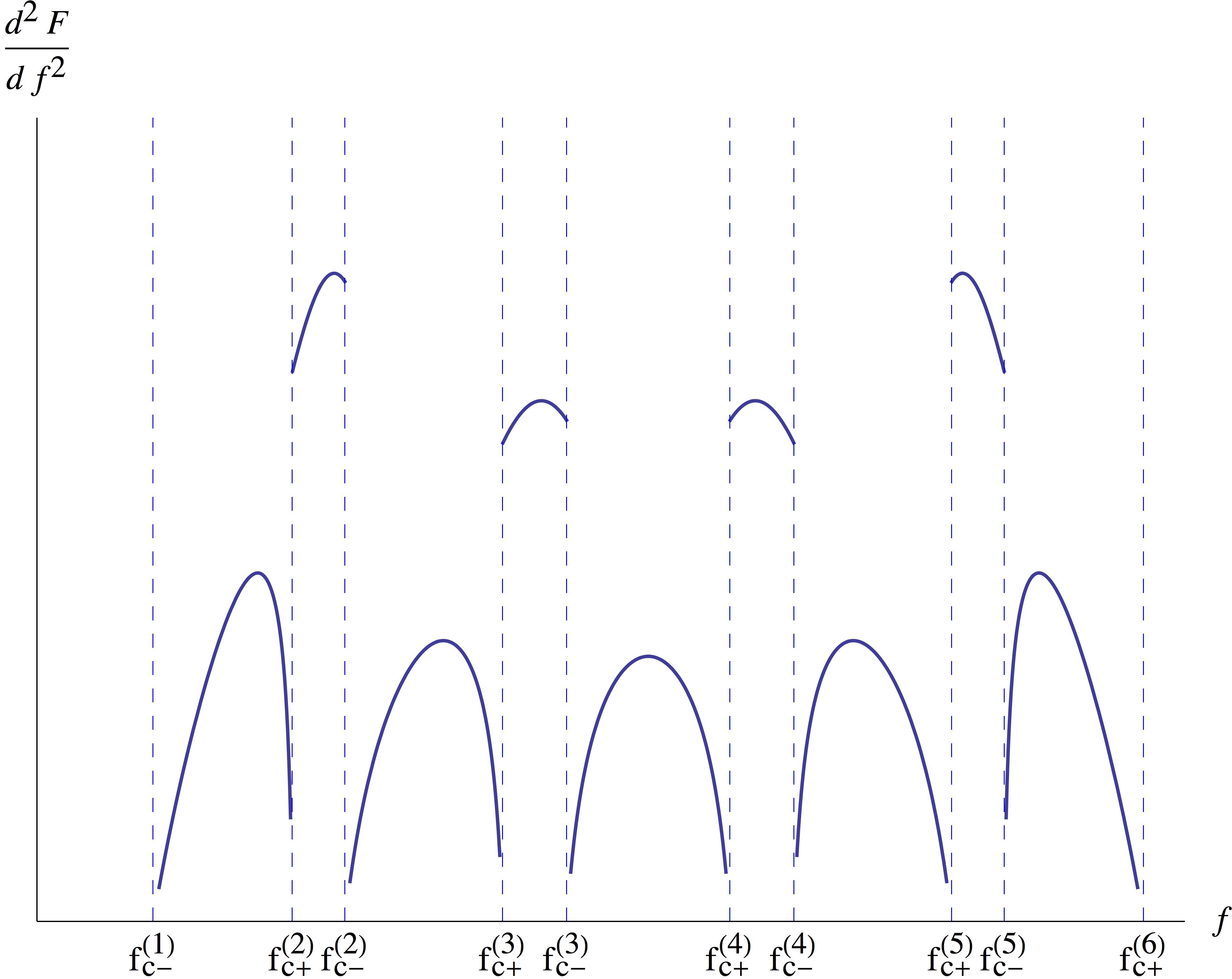}
   \label{fig:subfig2}
 }
\caption{\small (a) The eigenvalue density, for $n=6$, along with the average Wigner distribution shown as the dashed line. (b) The second derivative of the free energy, for the same set of parameters.}
\end{center}
\end{figure}

The width of the eigenvalue distribution $2\mu $ grows with $\lambda $, and the number of cusps accordingly multiplies. At strong coupling the density gets a rather complicated short-scale structure with a large number of cusps, while the average density is very simple and is described by the Wigner distribution, as shown in fig.~\ref{fig:subfig1}. To the first two orders of the strong-coupling expansion \cite{Chen:2014vka,Zarembo:2014ooa}, 
\begin{eqnarray}\label{bulk-density}
 \rho  (x)&\simeq& \frac{2}{\pi \mu ^2}\,\sqrt{\mu ^2-x^2}
\nonumber \\
&&
 +\frac{1}{\pi }\sqrt{\frac{M}{2\mu ^5}}\left[
 \left(\mu -x\right)\zeta \left(\frac{1}{2}\,,\left\{\frac{\mu +x}{M}\right\}\right)
 +\left(\mu +x\right)\zeta \left(\frac{1}{2}\,,\left\{\frac{\mu -x}{M}\right\}\right)
 \right].
\end{eqnarray}
The width of the eigenvalue distribution, to the same accuracy, is given by
\begin{equation}
 \mu \simeq \frac{\sqrt{\lambda }M}{2\pi }\,,
\end{equation}
but it is more convenient to regard $M$ and $\mu $ as independent variables (rather than $M$ and $\lambda $).

Taking into account that the zeta function has the square-root branch point at zero: $\zeta (1/2,z)\simeq 1/\sqrt{z}$, we find that the singular points (\ref{SingN2*}) are left or right cusps:
\begin{equation}
 \rho (x)\stackrel{x\rightarrow x^{(k)}_{c_\pm}}{=}
\begin{cases}
 \frac{C_\pm^{(k)}}{\sqrt{|x-x^{(k)}_{c_\pm}|}} & {\rm at}~x\rightarrow x_{c_\pm}\mp 0
\\
  \rho^{(k)} _0 & {\rm at}~x\rightarrow x_{c_\pm}\pm 0,
\end{cases}
\end{equation}
where
\begin{equation}
 \rho _0^{(k)}=\frac{2M\sqrt{k\left(n-k\right)}}{\pi \mu ^2}\,,
\end{equation}
and
\begin{equation}
 C_+^{(k)}=\frac{M^2(n-k)}{\sqrt{2\pi ^2\mu ^5}}\,,\qquad
  C_-^{(k)}=\frac{M^2k}{\sqrt{2\pi ^2\mu ^5}}\,. 
\end{equation}
Notice that $\rho _0^{(k)}\sim 1/\mu \gg C_\pm^{(k)}/\sqrt{\mu }\sim M/\mu ^2$, so the singularities are parametrically weak.

The oscillating structure in the density (\ref{bulk-density}) is a small correction on top of the regular Wigner distribution, and in addition the irregular part of the eigenvalue density integrates to zero upon averaging over sufficiently large interval, up to $O(M^{3/2}/\mu ^{3/2})$ which is beyond the accuracy of (\ref{bulk-density}). Therefore, to the leading order, the expectation value of the anti-symmetric Wilson loop is given by the formulas for the Gaussian matrix model \cite{Yamaguchi:2006tq}:
\begin{equation}
 F=\frac{2\mu }{3\pi }\,\sin^3\theta ,
\end{equation}
where $\cos\theta=\nu /\mu  $, with $\nu $ from (\ref{statmech}) related to the representation variable $f$ by
\begin{equation}\label{f-theta}
 \pi f=\theta -\frac{1}{2}\,\sin 2\theta .
\end{equation}

Taking into account the next order introduces the oscillating peak structure in the eigenvalue density and leads to the phase transitions in the Wilson loop expectation value.
According to (\ref{SingN2*}), the phase transitions happen at $f=f_{c_\pm}^{(k)}$, where the critical representation labels $f_{c_+}^{(k)}$ are given by (\ref{f-theta}) with
\begin{equation}
 \cos\theta _{c_+}^{(k)}=1-\frac{2k}{n+\Delta }\,,\qquad k=1,\ldots  ,n
\end{equation}
and
\begin{equation}
 f_{c_-}^{(k)}=1-f_{c_+}^{(n-k+1)}.
\end{equation}
Although $n\gg \Delta $, it is important to keep $\Delta $ in this equation to break the degeneracy under $\theta \rightarrow \pi -\theta $ and to get right the positions of the critical points.

The second derivative of the free energy experiences a jump (\ref{jump}) at the $k$-th critical point of an amplitude
\begin{equation}
  \left.\frac{d^2F}{df^2}\right|_{f^{(k)}_{c_\pm}-0}^{f^{(k)}_{c_\pm}+0}=\pm\frac{\pi \mu ^2}{2M\sqrt{k(n-k)}}\,.
\end{equation}
Since $d^2F/df^2$ at each critical point jumps to zero, its discontinuity is of the same order $\mathcal{O}(\mu )$ as its average value. This leads to the structure displayed in fig.~\ref{fig:subfig2}.

The above calculations apply to $f=O(1)$. For parametrically smaller $f=O(\lambda ^{-3/4})$ the endpoint region of the eigenvalue distribution, where (\ref{bulk-density}) is no longer accurate, becomes more important. The phase transitions  for $f=O(\lambda ^{-3/4})$ were analyzed in detail in \cite{Chen-Lin:2015dfa}. In both cases $\lambda $ is assumed to be large, and these results are potentially relevant for the holographic description \cite{Pilch:2000ue,Buchel:2000cn} of the $\mathcal{N}=2^*$ theory.

\section{Conclusions}

Wilson loops in large antisymmetric representations undergo phase transitions  which mirror quantum phase transitions in the underlying gauge theories. Both are caused by features in the eigenvalue density, delta functions or cusps, known to occur in localization matrix models for $\mathcal{N}=2$ gauge theories. In four dimensions only these singularities can arise. It would be interesting to extend the analysis  to localization matrix models in other dimensions where different types of singularities may occur. 

Holographically, anti-symmetric Wilson loops of rank $k\sim N$ correspond to probe D5-branes in the dual geometry \cite{Yamaguchi:2006tq}. The  classical solutions for D5-branes are known only for $AdS_5\times S^5$ \cite{Yamaguchi:2006tq}. It would be extremely interesting to construct D5-brane configurations for the backgrounds that are dual to massive models with phase transitions, for instance in the Pilch-Warner background \cite{Pilch:2000ue} dual to $\mathcal{N}=2^*$ SYM. So far only D3-brane solutions, which correspond to Wilson loops in large symmetric representations, have been constructed for this background \cite{Chen-Lin:2015xlh}. The latter are affected by singularities in the eigenvalue density to much lesser degree compared to antisymmetric representations and are not very sensitive to the phase transitions.

\section*{Acknowledgements}

J.G.R. acknowledges financial support from a MINECO grant FPA2016-76005-C2-1-P
and  MDM-2014-0369 of ICCUB (Unidad de Excelencia `Mar\'ia de Maeztu').
The work of K.~Z.  was supported by the grant ``Exact Results in Gauge and String Theories'' from the Knut and Alice Wallenberg foundation, by the ERC advanced grant No 341222, by the Swedish Research Council (VR) grant
2013-4329 and by RFBR grant 15-01-99504.


\begin{appendix}

\def\KK{{\cal K}}

\section{SQCD with several mass scales}\label{scales-appx}

Consider ${\cal N}=2$ supersymmetric $SU(N)$ QCD on ${\mathbb S}^4$, with $n_{f}$ pairs of fundamental hypermultiplets of mass $(m,-m)$, and $N_{f}$
pairs of fundamental hypermultiplets of mass $(M,-M)$, with $M > m$. The case $m=M$ is the theory studied in section \ref{sectionSQCD}.

The partition function computed by localization is given by
\be\label{NfQCD-partfunc}
Z^{\rm SQCD}_{N_f} =  \int d^{N-1}a\, \,\frac{{\rm e}\,^{2(N-N_f-n_f )\left(\ln\Lambda R +\gamma +1\right)\sum\limits_{i}  a_i ^2R^2}\prod_{i<j}^{}\left(a_i-a_j\right)^2 H^2(a_i-a_j)}{\prod_{i}^{}H^{N_f}(a_i+M) H^{N_f}(a_i-M)H^{n_f}(a_i+m) H^{N_f}(a_i-m)   }
\,\,,
\ee
where $R$ is the radius of   ${\mathbb S}^4$ and  $\Lambda $ is  the dynamically generated scale
\begin{equation}
 \Lambda R=\,{\rm e}\,^{-\frac{4\pi ^2}{\lambda \left(1-\zeta \right)}},
\end{equation}
with Veneziano parameters
\be
\zeta=\zeta_M+\zeta_m\ ,\qquad \zeta_M=\frac{N_f}{N}\ ,\qquad \zeta_m=\frac{n_f}{N}\ .
\ee
We shall consider $\zeta_M+\zeta_m<1$, in which case the theory is asymptotically free.
The partition function  is thus expressed in terms of
a finite dimensional integral, which is still difficult to compute. 
 Here, as section \ref{sectionSQCD} (see also \cite{Russo:2013kea,Russo:2013sba}), we
shall compute this integral in the planar, large $N$ limit at fixed $\lambda $, $\zeta_m,\ \zeta_M$, by the saddle-point method. 

Let us first summarize the physical picture of the $m=M$ case considered in section \ref{sectionSQCD}. The theory has two phases, weak coupling $2\Lambda<M$ and strong coupling $2\Lambda>M$, separated by a third-order phase transition at $2\Lambda_c =M$.
In this new theory with three mass scales $\Lambda, \ M, \ m$, the question we would like to address is whether there are
more phase transitions and at which value of $\Lambda $ they occur. A naive guess would be
that a new phase transition should occur at $\Lambda \sim m$. However, when $\Lambda\ll M$, the hypermultiplets of mass $M$ can be integrated out and the new dynamical scale of the theory is not $\Lambda $: it is replaced by a new effective scale that we shall compute. For example, in the theory of  \cite{Russo:2013kea,Russo:2013sba} with a single mass scale, at  $\Lambda\ll M$ the theory flows to pure ${\cal N}=2$ SYM with 
$\Lambda_{\rm eff} =\Lambda^{1-\zeta_M} M^{\zeta_M}$.
In the new theory with two mass scales $M, \ m$, the next phase transition is expected not when $\Lambda
= O(m)$, but, perhaps,  when the new dynamical scale of the theory (left behind after the hypermultiplets of mass $M$ have been integrated out) is of order $m$. The present exact calculation will 
determine which are the effective scales in each phase, and at which precise coupling $\Lambda $ the different phase transitions occur.

In the large $N$ limit, the saddle-point equation becomes
the following integral equation:
\bea
&& 
2\strokedint_{-\mu }^{\mu } dy \rho(y) \left(\frac{1}{x-y} -\KK(x-y)\right)
= -4\left(1-\zeta_M-\zeta_m \right) \left(\ln \Lambda+\gamma +1\right)x 
\nonumber\\
&& - \zeta_M   \KK( x + M) - \zeta_M   \KK( x - M)  - \zeta_m   \KK( x + m) - \zeta_m   \KK( x - m)\ .
\label{fresa}
\eea
where $\KK = x(\psi (1+ix) +\psi(1-ix)+2\gamma)$,
$\psi $ represents as usual the logarithmic derivative of the $\Gamma $-function and $\gamma$ 
is the Euler constant, $\gamma=-\psi(1)$ 
(see \cite{Russo:2013kea} for the
relation of $\KK $ to the Barnes G-function and other properties).
Here we have set $R=1$ for the sake of clarity.

The  model depends on
three parameters $\Lambda R, MR$ and $mR$.
In the decompactification limit, these parameters go to infinity and the saddle-point equations simplify.
In this limit, the eigenvalue distribution extends to large eigenvalues and one can use the approximation
$\KK = x\ln x^2 +2\gamma x + O(x^{-1})$.
The first  derivative of the saddle-point equations is given by
\be\label{auxiliary-SQCD}
2\int_{-\mu }^{\mu } dy\,\rho(y)\ln \frac{\left(x-y\right)^2}{\Lambda ^2}
=  \ln\frac{\left(x^2-M^2\right)^{2\zeta_M} \left(x^2-m^2\right)^{2\zeta_m}}{\Lambda ^{4\zeta}}\,.
\ee
Differentiating once more we obtain a singular integral equation that can be easily solved:
\be
\label{mfre}
2\strokedint_{-\mu }^{\mu } dy\,\,  \frac{\rho(y)}{x-y}
=   \frac{\zeta_M }{x+M} +\frac{\zeta_M }{x-M}+ \frac{\zeta_m }{x+m} +\frac{\zeta_m }{x-m}\,.
\ee
Just like in the model of section \ref{sectionSQCD},
corresponding to the $m=M$ case (or to the $\zeta_m=0$ case),
the solution to this equation (\ref{mfre}) depends on whether $x=\pm m$ or $x=\pm M$ lie inside or outside the cut $(-\mu,\mu)$.
The system has therefore three phases:

\begin{itemize}

\item i) Strong coupling, $\mu>M>m$. The poles at $x=\pm M$ and $x=\pm m$ lie inside the cut. The eigenvalue density $\rho(x)$ has delta function terms $\frac12 \zeta_M \delta(x\pm M)$ and  $\frac12 \zeta_m \delta(x\pm m)$. 

\item ii) Intermediate coupling, $M>\mu>m$. The poles at $x = m$ and at  $x = -m$ lie on the cut. The eigenvalue density $\rho(x)$ now contains delta function terms   $\frac12 \zeta_m \delta(x\pm m)$. 

\item iii) Weak coupling phase, $M>m>\mu $. All poles lie outside the eigenvalue distribution.  

\end{itemize}

In what follows we describe the analytic solution in each phase.

\subsubsection*{Strong-coupling phase ($\mu>M>m$)} 

When $\mu>M>m$, the  eigenvalue density that solves  (\ref{mfre}) is  given by
\be
\label{noj}
\rho (x) = \frac{1-\zeta }{\pi \sqrt{\mu^2-x^2}} + \frac{\zeta_M }{2}\,\delta(x+M)+\frac{\zeta_M }{2}\,\delta(x-M)
+ \frac{\zeta_m }{2}\,\delta(x+m)+\frac{\zeta_m }{2}\,\delta(x-m).
\ee
The parameter $\mu $ is determined in terms of $M,\ m $ and $\Lambda $  by
 substituting the solution into (\ref{auxiliary-SQCD}). This gives
\begin{equation}\label{mu=2Lambda}
 \mu =2\Lambda .
\end{equation}
Note that in this phase the endpoint of the eigenvalue distribution is therefore independent of the hypermultiplet masses.

The solution (\ref{noj}) holds
as long as $M<\mu $. As $\Lambda $ is gradually decreased, there is a critical point $\Lambda_{c1}$
where  $\mu = M$, which thus occurs at
\begin{equation}
 \Lambda_{c1}=\frac{M}{2} .
\end{equation}
For $\Lambda <\Lambda_{c1}$, the delta-functions $\delta(x\pm M)$ lie outside the interval $[-\mu ,\mu ]$ and  the density (\ref{noj}) does not solve (\ref{mfre}) anymore.  
This happens just like in the $m=M$ case discussed
in section \ref{sectionSQCD}.

\subsubsection*{Intermediate-coupling phase ($M>\mu>m$)}

 In the regime $M>\mu>m$, the solution is given by
\begin{equation}
\label{rhointer}
 \rho (x)=\frac{1-\zeta}{\pi \sqrt{\mu ^2-x^2}}
+ \frac{\zeta_M M\sqrt{M^2-\mu ^2}}{\pi \sqrt{\mu ^2-x^2}\left(M^2-x^2\right) }+ \frac{\zeta_m }{2}\,\delta(x+m)+\frac{\zeta_m }{2}\,\delta(x-m)
 .
\end{equation}
Substituting into (\ref{auxiliary-SQCD}) we find the following transcendental equation for $\mu $:

\begin{equation}
\label{nilo}
\left(1-\zeta \right)\ln\frac{\mu }{2M}+\zeta_M \ln\frac{\mu }{M+\sqrt{M^2-\mu ^2}}
=\left(1-\zeta \right)\ln\frac{\Lambda }{M}\,.
\end{equation}

In the particular case $\zeta_m=0$, corresponding to $n_f=0$ and giving $\zeta_M=\zeta$, the resulting equation was studied in \cite{Russo:2013kea,Russo:2013sba}.
Like in this case,  the solution can be  expressed in a parametric form:
\begin{eqnarray}\label{muuu}
\mu &=&M\sqrt{1-u^2},
\\
  \left(\frac{2\Lambda }{M}\right)^{2-2\zeta_m-2\zeta_M }&=&\left(1+u\right)^{1-\zeta_m-2\zeta_M }\left(1-u\right)^{1-\zeta_m}.
 \label{udefinition}
\end{eqnarray}
Here we can see the emergence of the first relevant effective scale. When $M\gg\Lambda $, the hypermultiplets can be integrated out. From the above equations we obtain
\be
\label{lambeff}
\left( \frac{\mu}{2} \right)^{1-\zeta_m} \approx M^{\zeta_M} \Lambda^{1-\zeta}\ ,\qquad M\gg\Lambda \ .
\ee
The theory left behind should  be SQCD with $n_f$ flavors and dynamical QCD scale $\Lambda_{\rm eff1}\sim \mu$.

Note that (\ref{nilo}), (\ref{udefinition}) simplify in the particular case $\zeta_m+2\zeta_M =1$. Then
we can explicitly solve for $\mu $
\be
\mu = 2\sqrt{\Lambda (M-\Lambda)}\ .
\ee

There is another interesting case, $\zeta_M =\frac34 (1-\zeta_m)$, where the equation (\ref{nilo}) simplifies and can be solved explicitly. 
Then we find
\be 
\mu =\frac{\Lambda}{\sqrt{2}} \left(\sqrt{1+\frac{4M}{\Lambda}}-1\right)^{\frac32 } \ .
\ee
For $\Lambda\ll M$, one has
\be
\mu \approx 2 M^{\frac34}\Lambda^{\frac14}  =\Lambda_{\rm eff1}\ ,
\ee
in agreement with (\ref{lambeff}). 


\subsubsection*{Weak-coupling phase ($M>m>\mu $)}

As $\Lambda $ is further decreased, there is a critical point $\Lambda_{c2}$ where $\mu=m$.
For lower values of $\Lambda$, 
 the delta-functions $\delta(x\pm m)$ move outside the eigenvalue region $[-\mu ,\mu ]$ and  the density (\ref{rhointer}) does no longer solve (\ref{mfre}).  
 In this regime $M>m>\mu $ we find the solution

\begin{equation}
 \rho (x)=\frac{1-\zeta}{\pi \sqrt{\mu ^2-x^2}}
 +\frac{\zeta_M M\sqrt{M^2-\mu ^2}}{\pi \sqrt{\mu ^2-x^2}\left(M^2-x^2\right)}
+\frac{\zeta_m m\sqrt{m^2-\mu ^2}}{\pi \sqrt{\mu ^2-x^2}\left(m^2-x^2\right)}\ .
\end{equation}

Substituting into the integrated form of the saddle-point equation (\ref{auxiliary-SQCD}),  we
now find the following transcendental equation for $\mu $:

\begin{equation}
\label{rio}
\left(1-\zeta \right)\ln\frac{\mu }{2}+\zeta_M \ln\frac{\mu }{M+\sqrt{M^2-\mu ^2}}
+\zeta_m \ln\frac{\mu }{m+\sqrt{m^2-\mu ^2}}=\left(1-\zeta \right)\ln \Lambda \,.
\end{equation}
This defines $\mu= \Lambda  f(m/\Lambda,M/\Lambda)$.
In particular, consider the case $\Lambda\ll m,M$. In this case, all hypermultiplets can be integrated out
and the theory should be pure ${\cal N}=2$ SYM with an effective dynamical scale.
The question is which  the effective dynamical scale is.
From (\ref{rio}), in this limit we find
\be
\mu \approx 2 m^{\zeta_m} M^{\zeta_M} \Lambda^{1-\zeta_M-\zeta_m}\equiv \Lambda_{\rm eff2}\ .
\ee
This can also be written in the form
\be
\Lambda_{\rm eff2} = \bar M \left( \frac{\Lambda}{\bar M} \right)^{1-\zeta}\ ,\qquad \bar M^{\zeta_m+\zeta_M} \equiv m^{\zeta_m} M^{\zeta_M}\ .
\ee
Concerning the interpretation of $\Lambda_{\rm eff2} $ as the effective dynamical scale, one can make the following consistency check. We note that 
$\Lambda_{\rm eff2} \approx \Lambda_{\rm eff1}^{1-\zeta_m} m^{\zeta_m}$. This is indeed the dynamical scale that one should expect when $n_f=\zeta_m N$ flavors are integrated out starting from a theory with scale $\Lambda_{\rm eff1}$.

Let us now determine at which value of the original coupling $\Lambda_{c2} $  the  phase transition from the intermediate to the weak-coupling phase occur.
This is computed by setting $\mu =m$ in (\ref{rio}) (or, equivalently, in (\ref{nilo})).
This gives
\be
\left(1-\zeta \right)\ln 2\Lambda_{c2} =\left(1-\zeta_m \right)\ln m - \zeta_M \ln(M+\sqrt{M^2-m^2})
\ .
\end{equation}
Note that $M+\sqrt{M^2-m^2}=\alpha M$, where $1<\alpha<2$. 
Therefore, ignoring a factor of order 1, we can substitute $M+\sqrt{M^2-m^2}$ by $M$.
Then
\be
\Lambda_{c2}^{1-\zeta}\approx  m^{1-\zeta_m} M^{-\zeta_M}\ ,
\ee
or
\be
\Lambda_{c2} \approx m \left(\frac{m}{M}\right)^{\frac{\zeta_M}{1-\zeta}}\ .
\ee
This last expression shows, in particular, that $\Lambda_{c2}\ll m$ if $m\ll M$.

Substituting  $\Lambda=\Lambda_{c2}$ into $\Lambda_{\rm eff2}$, we find that the new phase transition 
 occurs at
\be
\Lambda_{\rm eff2} \approx m\ .
\ee
Note also that, for this $\Lambda $, one also has $\Lambda_{\rm eff1} \approx \Lambda_{\rm eff2} $.

Following \cite{Russo:2013kea,Russo:2013sba}, it is easy to show that the two first transitions occurring at $\Lambda_{c1}, \ \Lambda_{c2}$ are third order.
To show this, one computes the different derivatives of the free energy at each phase
using
\be
\Lambda \frac{\partial F}{\partial \Lambda } = -2(1-\zeta) \int_{-\mu}^{\mu} dx \rho(x) x^ 2\ .
\ee
[Here $F=-\ln Z$ is the standard free energy of the theory --it should not be confused with the $F(f)$ appearing in (\ref{wilAA}), which is related to the free energy of the effective Fermi distribution]. Likewise, one can show that
the Wilson loop in the fundamental representation is discontinuous in its first derivatives: recall that $\ln W_f\sim 2\pi \mu $, therefore $\ln W_f$ inherits the first-derivative discontinuities of $\mu$ at $\Lambda_{c1}, \ \Lambda_{c2}$.

Summarizing, starting from strong coupling $\Lambda\gg M,\ m$, as $\Lambda $ is gradually decreased, the theory undergoes a first phase transition 
when $\Lambda =M/2$. As $\Lambda $ is further reduced, one finds that nothing happens when $\Lambda =O(m)$; the theory has a smooth behavior with the coupling $\Lambda $ at this scale.  The new phase transition to the weak-coupling phase occurs at a lower scale 
$\Lambda_{c2}$ (which becomes
much lower if $m\ll M$). Once the hypermultiplets of mass $M$ have been integrated out,
the theory left behind has a new dynamical scale  $\Lambda_{\rm eff1} $. As one may expect, if a new phase transition occurs, this one will take place when $\Lambda_{\rm eff1} $ is of order $m$ and this is indeed what happens.
It is reassuring that all field theory expectations
are realized in a precise way through the exact large $N$ solution of the system.

\section{Useful integrals}

The formulas for the integrals used in section 2 can be derived by residue integration.
One chooses a contour surrounding the cut from $(-\mu,\mu)$ and evaluates the residue of the poles outside the cut (including possible poles at
infinity) by the change of variable $z=1/x$.
One finds
\be
\int_{-\mu}^\mu dx \frac{1}{\pi\sqrt{\mu^2 -x^2}} =1
\ee
\be
\int_{-\mu}^\mu dx \frac{1}{\pi\sqrt{\mu^2 -x^2}}\frac{1}{x-y} =0
\ee
\be
\int_{-\mu}^\mu dx \frac{1}{\pi\sqrt{\mu^2 -x^2}}\ \ln x^2=\ln \frac{\mu^2}{4}
\ee
For $\mu<M$, by residue integration, one finds
\be
M\sqrt{M^2-\mu^2}\int_{-\mu}^\mu dx \frac{1}{\pi\sqrt{\mu^2 -x^2}}\frac{1}{M^2-x^2}=1\ \ ,
\ee
\be
M\sqrt{M^2-\mu^2}\int_{-\mu}^\mu dx \frac{1}{\pi\sqrt{\mu^2 -x^2}}\frac{1}{M^2-x^2}\frac{1}{x-y}=\frac{1}{x+M}+\frac{1}{x-M}\ \ ,
\ee
\be
M\sqrt{M^2-\mu^2}\int_{-\mu}^\mu dx \frac{1}{\pi\sqrt{\mu^2 -x^2}}\frac{1}{M^2-x^2}\ \ln \frac{x^2}{M^2}=2 \ln \frac{\mu}{M+\sqrt{M^2-\mu^2}}\ \ .
\ee

\end{appendix}

\bibliographystyle{nb}

\end{document}